\documentclass[showpacs,aps,pra,amsmath,amssymb,twocolumn]{revtex4-1}
\usepackage{graphicx}
\usepackage{dcolumn}
\usepackage{bm,color}
\usepackage{txfonts}
\usepackage{amsmath,epsfig}
\usepackage{epstopdf}
\usepackage{times}

\newcommand{\eq}{Eq.~}
\newcommand{\eqs}{Eqs.~}
\newcommand{\fig}{Fig.~}
\newcommand{\figs}{Figs.~}
\newcommand{\cf} {cf.~}
\newcommand{\ug} {\!=\!}
\newcommand{\tens} {\!\otimes\!}
\newcommand{\piu} {\!+\!}
\newcommand{\meno} {\!-\!}
\newcommand{\ie} {i.e.~}

\newcommand{\vs} {vs.~}

\newcommand{\boldsigma} {\mbox{\boldmath$\sigma$}}

\begin{document}

\author{Francesco Ciccarello}
\author{Vittorio Giovannetti}

\affiliation{NEST, Scuola Normale Superiore and Istituto Nanoscienze-CNR, Piazza dei Cavalieri 7, I-56126 Pisa, Italy}

\title{Local-channel-induced rise of quantum correlations in continuous-variable systems}
 
\date{\today}
\begin{abstract}
It was recently discovered that the quantum correlations of a pair of disentangled qubits, as measured by the quantum discord, can increase solely because of their interaction with a {\it local} dissipative bath. Here, we show that a similar phenomenon can occur in continuous-variable bipartite systems. To this aim, we consider a class of two-mode squeezed thermal states and study the behavior of Gaussian quantum discord under various {\it local} Markovian non-unitary channels. While these in general cause a monotonic drop of quantum correlations, an initial rise can take place  with a thermal-noise channel.
\end{abstract}
\maketitle

\section{Introduction}

Within the remit of quantum information processing (QIP) theory \cite{nc} and beyond, a paramount topic is the study of quantum correlations (QCs). Until recently, the emergence of QCs has been regularly highlighted in connection with non-separable states of multipartite quantum systems \cite{horo1}. The violation of the celebrated Bell inequalities is a typical signature of the extra amount of correlations that quantum systems can possess besides those of a purely classical nature \cite{nc, horo1}. Non-separability, \ie entanglement, which relies on the superposition principle, has been in fact regarded as a necessary prerequisite in order for QCs to occur. In 2001, however, it was discovered \cite{seminal} a new way in which the superposition principle can entail an alternative type of non-classical correlations even in the {\it absence of any entanglement}. Typical instances are mixtures of pure separable states which are locally non-orthogonal, \ie indistinguishable \cite{talmor}. More rigorously, a state has non-classical correlations whenever its associated density operator cannot be diagonalized in a basis which is the tensor product of local orthonormal bases \cite{talmorsimil}. When this occurs the entire correlations' content cannot be retrieved by any local measurement at variance with the classical framework where this is always achievable.
Following such breakthrough, a growing interest has developed especially after the realization that this new QCs'~paradigm could be key to some known entanglement-free QIP schemes \cite{ent-free}.

Various measures have been proposed in the literature to detect such kind of QCs. A prominent one is {\it quantum discord} \cite{seminal} whose definition simply arises from the quantum generalization of two classically-equivalent versions of the mutual information (one being based on the conditional entropy). These are found to differ in the quantum framework, the corresponding discrepancy just being the quantum discord. This encompasses both QCs associated with entanglement and those that can be exhibited by separable states. The non-equivalence between such two forms of non-classical correlations arises when mixed states are addressed since as long as the state is pure the discord coincides with the entropy of entanglement. 

Significant motivations to pursue a deeper understanding of these problems are currently being provided by various investigations targeting non-unitary dynamics. While entanglement is known to be extremely fragile to environmental interactions, QCs -- as given by such extended notion -- are generally very robust and in some cases even fully insensitive to phase noise \cite{laura}. This is related to the fact that zero-discord states are a set of negligible measure within the entire Hilbert space \cite{subset}. 
Still in the framework of non-unitary dynamics, very recently it was found that the interaction with a {\it local and memoryless} bath can even create QCs initially fully absent \cite{ciccarello, mauro,bruss, gong}. Such effect, which is unattainable with entanglement \cite{horo1}, was shown to take place in particular for qubits prepared in a fully classical state and undergoing a local amplitude-damping channel. This describes the dissipative interaction with a local bath (such as the spontaneous emission of a two-level atom) \cite{nc}. Specifically, the behavior consists of an initial rise of QCs until a maximum is reached followed by a slow decay, the entanglement being zero throughout. The essential underlying mechanism is that such non-unitary process can map orthogonal states of a subsystem onto non-orthogonal ones. The state thus can no more be diagonalized in a tensor product of orthonormal bases, \ie it acquires QCs \cite{ciccarello, bruss}.

Most of the work carried out so far along these lines, though, tackled qubits. It is however natural to wonder how the above findings are generalized for quantum {\it continuous-variable} (CV) systems, which routinely occur in quantum optics. Such investigations are still in their infancy. Indeed, the first measure of generalized QCs for  Gaussian states of bipartite CV systems has been worked out only last year \cite{dattaadesso, paris}. Such quantity, called {\it Gaussian discord}, is basically defined in the same spirit of standard discord \cite{seminal} and likewise can be non-zero for separable states \cite{dattaadesso, paris}. Recently, another measure has been proposed \cite{adessoAMID}.
To date, only few studies \cite{paris, ruggero} have targeted the non-unitary dynamics of Gaussian discord in the presence of environmental interactions including one model of non-Markovian reservoir \cite{nota}. It was found that apart from the case of a common reservoir the non-unitary dynamics is detrimental to QCs \cite{paris, ruggero,torun}.

Our goal in this work is to analyze the behavior of Gaussian discord in the case of two CV modes under the most relevant known instances of local -- \ie single-mode -- Gaussian memoryless channels \cite{eisert, holevo,njp} each described by an associated completely positive quantum map. Our major motivation is to assess whether growth of QCs can take place for some local non-unitary channels as in the case of qubits \cite{ciccarello,mauro,bruss}. We will indeed find that in significant analogy with the qubits' framework this can occur with the {\it lossy} channel (more in general with low-temperature thermal-noise channels).

This paper is organized as follows. In Section \ref{discord2}, we briefly review the definition of Gaussian discord. In Section \ref{states}, we describe the class of squeezed thermal states, which is the one which will be our focus in this work. In Section \ref{channels}, we review each of the aforementioned Gaussian channels and the related main features that we will refer to. In Section \ref{effect}, we show the behavior of QCs under each of the considered channel for some paradigmatic initial states. In Section \ref{insight}, we provide a simple picture that allows to obtain insight into the effects presented in the previous section. Finally, in Section \ref{concl} we draw our conclusions.

\section{Quantum discord of Gaussian states: review}  \label{discord2} 

Given two systems $1$ and $2$ in a state $\rho$, the quantum mutual information is defined as
\begin{equation}\label{mutinf}
\mathcal{I}(\rho)\ug S(\rho_1)\piu S(\rho_2)-S(\rho)\,\,
\end{equation}
where $\rho_{1(2)}\ug{\rm Tr}_{2(1)}\rho$ is the reduced density operator describing the state of 1 (2) and $S(\sigma)\!=\!-{\rm Tr}\,(\sigma{\rm\, log} \sigma)$ is the Von Neumann entropy of an arbitrary state $\sigma$.
A local generalized measurement on $2$ can be specified by a complete set of positive-operator-valued projectors (POVM) $\{\Pi_k\}$, where $k$ indexes a possible outcome. If $k$ is recorded with probability $p_k\!=\!{\rm Tr}[\rho \openone \tens \Pi_k]$ the overall system collapses onto the (normalized) state $\rho_k\!=\!(\rho \openone \tens \Pi_k)/p_k$. The maximum of $S(\rho_1)\meno \sum_k p_k S(\rho_k)$, \ie the mismatch between the entropy of 1 and the average conditional entropy, reads 
\begin{equation}\label{j}
\mathcal{J}(\rho)\ug \max_{\{\Pi_k\}} \left[S(\rho_1)\meno \sum_k p_k S(\rho_k)\right]\,\,,
\end{equation}
and is taken over all possible POVM measurements {each described by the operator $\Pi_k$}. If 1 and 2 are CV modes and one restricts to generalized {\it Gaussian} measurements 
\cite{cirac} the Gaussian discord $\mathcal{D}^\leftarrow$ is defined as the discrepancy between (\ref{mutinf}) and (\ref{j}) \cite{paris,dattaadesso}
\begin{equation}\label{disc1}
\mathcal{D}^\leftarrow=\mathcal{I}-\mathcal{J}\,\,.
\end{equation}
The arrow reminds that measurements over mode 2 have been considered.

As is well-known, any two-mode zero-mean Gaussian state $\rho_{12}$ is fully specified by its covariance matrix $\sigma_{ij}\ug{\rm Tr\left[ \rho_{12}\left(R_{i}R_{j}\piu R_{j}R_{i}\right)\right]}$, where ${\bf R}\ug\left\{x_{1},p_{1},x_{2},p_{2}\right\}$ is the set of operators corresponding to the phase-space coordinates. Up to local symplectic, \ie unitary, operations the covariance matrix can be arranged in the form
\begin{equation} \label{sigma}
\mbox{\boldmath$\sigma$}\! =\!\left( \begin{array}{cc}
  A&           C \\
            C          &B \end{array} \right)\,\,,
 \end{equation}
where $A\ug{\rm diag}(a,a)$, $B\ug{\rm diag}(b,b)$ and $C\ug{\rm diag}(c_1,c_2)$. The determinants of such diagonal matrices $I_1\ug{\rm det}\,A$, $I_2\ug{\rm det}\,B$, $I_3\ug{\rm det}\,C$ along with $I_4\ug{\rm det}\mbox{\boldmath$\,\sigma$}$ are symplectic invariants, \ie they are invariant under local symplectic transformations. The quantum discord $\mathcal{D}^{\leftarrow}$ is a function of these symplectic invariants according to \cite{paris, dattaadesso}
\begin{equation}\label{disc}
\mathcal{D}^{\leftarrow}\ug h(\sqrt{I_2})-h(d_-)-h(d_+)+h\left(\frac{\sqrt{I_1}+2\sqrt{I_1I_2}+2I_3}{1+2\sqrt{I_2}}\right)\,\,,
\end{equation}
where $h(x)\ug(x\piu1/2)\ln(x\piu1/2)-(x\meno1/2)\ln(x\meno1/2)$ and
\begin{equation}\label{dpm}
d_{\pm}^2\ug\frac{\Delta\pm\sqrt{\Delta^2-4I_4}}{2}
\end{equation}
with $\Delta\ug I_1+I_2+2 I_3$. The discord in terms of measurements on mode 1 $\mathcal{D}^\rightarrow$ is obtained from (\ref{disc}) upon the replacement $I_1\leftrightarrow I_2$. 
In the remainder of this work, we will be solely interested in the discord (\ref{disc}), hence we will  drop the subscript henceforth and set $\mathcal{D}\ug \mathcal{D}^\leftarrow$.

\section{Two-mode squeezed-thermal states} \label{states}

In this work, we shall focus on the class of two-mode squeezed-thermal states (STSs) whose generic element reads
\begin{equation} \label{sts}
\rho\ug e^{r(a_1^\dagger a_2^\dagger-a_1a_2)} (\rho_1 \otimes \rho_2 )\; \left[e^{r(a_1^\dagger a_2^\dagger-a_1a_2)} \right]^\dagger\,\,,
\end{equation}
where each single-mode thermal state $\rho_i$ ($i\ug1,2$) is given by
\begin{equation}\label{thermal}
\rho_i\ug\sum_{n\ug0}^{\infty}\frac{N_i^n}{(1+N_i)^{n+1}}|n\rangle_i\langle n|\;,
\end{equation}
with $N_i$ the corresponding average number of photons.

The characteristic function $\chi(\lambda_1,\lambda_2)$ \cite{walls} is obtained from (\ref{sts})
as
\begin{equation}\label{chilambda}
\chi(\lambda_1,\lambda_2)\ug{\rm Tr}\left[\rho D(\lambda_1)D(\lambda_2)\right]\,\,,
\end{equation}
where $D(\lambda_i)\ug\exp(\lambda_i a_i^{\dagger}\meno\lambda^{*}a_i)$ is the displacement operator of the $i$th mode ($i\ug1,2$ with $a_i$ and $a_i^\dagger$ the usual ladder operators of the $i$th mode). The density operator is retrieved from the characteristic function as $\rho\ug\int d^2 \!\lambda_1 d^2\!\lambda_2\, \chi(\lambda_1,\lambda_2)D(\meno \lambda_1)D(\meno \lambda_2)/\pi^2$.
It is then straightforwardly checked that
\begin{equation}\label{cfsts}
\chi(\lambda_1,\lambda_2)\ug e^{-(N_1\piu 1/2)|\cosh r\lambda_1\meno \sinh r \lambda_2^*|^2}\,e^{-(N_2\piu 1/2)|\cosh r\lambda_2\meno  \sinh r  \lambda_1^*|^2}\,\,.
\end{equation}
As states (\ref{sts}) are Gaussian, the covariance matrix is related to the characteristic function according to \cite{ferraronotes}
\begin{equation}\label{chisigma}
\chi({\bf \Lambda})=\exp\left[ -\frac{{\bf \Lambda}^{\rm T}\boldsigma{\bf \Lambda}}{2}\right]\,\,,
\end{equation}
where ${\bf \Lambda\ug(\alpha_1,\beta_1,\alpha_2,\beta_2)}^{\rm T}$ and we have decomposed each complex variable $\lambda_j$ ($j\ug1,2$) as $\lambda_j\ug(\alpha_j\piu i \,\beta_j)/\sqrt{2}$. Upon comparison between \eqs (\ref{cfsts}) and (\ref{chisigma}) and in the light of (\ref{sigma}), the diagonal matrix elements defining $A$, $B$ and $C$ [\cf\eq(\ref{sigma})] are obtained as
\begin{eqnarray}
a&=&(1\piu N_r)N_1\piu N_r N_2\piu N_r\piu1/2\,\,,\label{abc1}\\
b&=&N_r N_1\piu (1\piu N_r) N_2\piu N_r\piu1/2\,\,,\label{abc2}\\
c_1&=&-c_2=-(1\piu N_1\piu N_2)\sqrt{ N_r (1\piu N_r)}\label{abc3}\,\,,
\end{eqnarray}
where we have set $N_r\ug(\sinh r)^2$.

\section{Local Gaussian channels} \label{channels}
In this Section, we review the salient features of the local one-mode Gaussian channels whose effect on the quantum correlations of two-mode states we aim to scrutinize. Throughout, we will assume that each of such channels acts on mode 2 only. 
The channels are Gaussian in that each of these maps Gaussian states into Gaussian states, hence allowing for use of Gaussian discord \cite{paris,dattaadesso} to shed light onto the QCs' behavior.
Specifically, we address the thermal-noise channel, which reduces to the lossy channel in the limit of vanishing reservoir temperature, the amplifier channel and the classical-noise channel \cite{eisert}.

\subsection{Thermal-noise channel}

This channel describes the dissipative interaction of a single-mode CV system with an environment at thermal equilibrium. Indeed, its associated map is routinely worked out by assuming that a single-mode environment in a thermal state specified by the average photon number ${N}$ is mixed by a beam splitter with the single-mode system. The state of the latter, and hence the map describing the channel, can then be obtained by simply tracing out the environmental degree of freedom. The channel quantum efficiency is measured by the parameter $\eta$ such that $0\!\le\!\eta\!\le\!1$ (in the above model $\eta$ is the transmissivity of the beam splitter).

Under the thermal-noise channel of efficiency $\eta$ and environment's average photon number $N$, the characteristic function $\chi(\lambda_1,\lambda_2)$ transforms into $\chi'(\lambda_1,\lambda_2)$ according to \cite{pra-lossy}
\begin{equation}\label{chitn}x
\chi'(\lambda_1,\lambda_2)\ug\chi(\lambda_1,\!\sqrt{\eta}\lambda_2)\,e^{{-(1-\eta)(N+1/2)|\lambda_2|^2}}\,\,\,\,\,\,\,\,(0\!\le\!\eta\!\le\!1)\,.
\end{equation}
For $N\ug0$ (zero-temperature environment) \eq(\ref{chitn}) reduces to the case of a {\it lossy channel} \cite{eisert} acting on mode 2.

By substituting (\ref{cfsts}) in \eq(\ref{chisigma}) and comparing this with \eq(\ref{chitn}) it is straightforwardly found that the covariance matrix keeps the same structure as in \eq(\ref{sigma}) with $A\ug{\rm diag}(a',a')$, $B\ug{\rm diag}(b',b')$ and $C\ug{\rm diag}(c_1',c_2')$. The new diagonal matrix elements $\{a',b', c'\}$ are related to the input ones [\cf \eqs(\ref{abc1})-(\ref{abc3})] according to
\begin{eqnarray}\label{abctnc}
a'&=&a\,\,,\\
b'&=&\eta b+(1-\eta)\left(N\piu\frac{1}{2}\right)\,\,,\\
c'_1&=&-c'_2=\sqrt{\eta}\, c_1\,\,.
\end{eqnarray}

\subsection{Amplifier channel}
This channel shares features similar to the thermal-noise channel but with the essential difference that it brings about an intensity amplification instead of a damping. Its corresponding map changes the characteristic function according to \cite{amplifier}
\begin{equation}\label{chiac}
\chi'(\lambda_1,\lambda_2)\ug\chi(\lambda_1,\!\sqrt{k}\lambda_2)\,e^{{-(k-1)(N+1/2)|\lambda_2|^2}}\,\,\,\,\,\,\,\,(k\!\ge\!1)\,.
\end{equation}
where $k\!\ge\!1$ is a gain parameter. In the limit $k\ug1$ the channel reduces to the identity operator.
By proceeding analogously to the case of the thermal-noise channel we find that the covariance matrix has again the same form as in \eq(\ref{sigma}) with the diagonal matrix elements now given by
\begin{eqnarray}\label{abcac}
a'&=&a\,\,,\\
b'&=&k b+(k-1)\left(N\piu\frac{1}{2}\right)\,\,,\\
c'_1&=&-c'_2=\sqrt{k}\, c_1\,\,.
\end{eqnarray}

\subsection{Classical-noise channel}

This channel arises when classical Gaussian noise is superimposed on the single-mode system \cite{cnchannel}. The parameter on which it depends is the number of injected noise photons $n\!\ge\!0$ in a way that the corresponding map becomes the identity in the limit $n\ug0$.
The characteristic function is transformed under this channel according to
\begin{equation}\label{chicn}
\chi'(\lambda_1,\lambda_2)\ug\chi(\lambda_1,\lambda_2)\;e^{-n|\lambda_2|^2}\,\,\,\,\,(n\!\ge\!0)\,\,.
\end{equation}
By comparing this with \eqs(\ref{cfsts}) and (\ref{chisigma}) it turns out that the covariance matrix has the form (\ref{sigma}) with diagonal matrix elements
\begin{eqnarray}\label{abccn}
a'&=&a\,\,,\\
b'&=&b+n\,\,,\\
c'_1&=&-c'_2=c_1\,\,.
\end{eqnarray}

\section{Behavior of quantum correlations} \label{effect}

Our aim in this Section is to present some typical behaviors of Gaussian discord when states {of the family} (\ref{sts}) are subject to the local Gaussian channels introduced in the previous Section. Unfortunately, a general analysis is quite demanding owing to the complicated functional form of (\ref{disc}). Nonetheless, as will become clear, one can obtain significant insight by addressing some paradigmatic instances.
\begin{figure*}
 \includegraphics[width=0.99\textwidth]{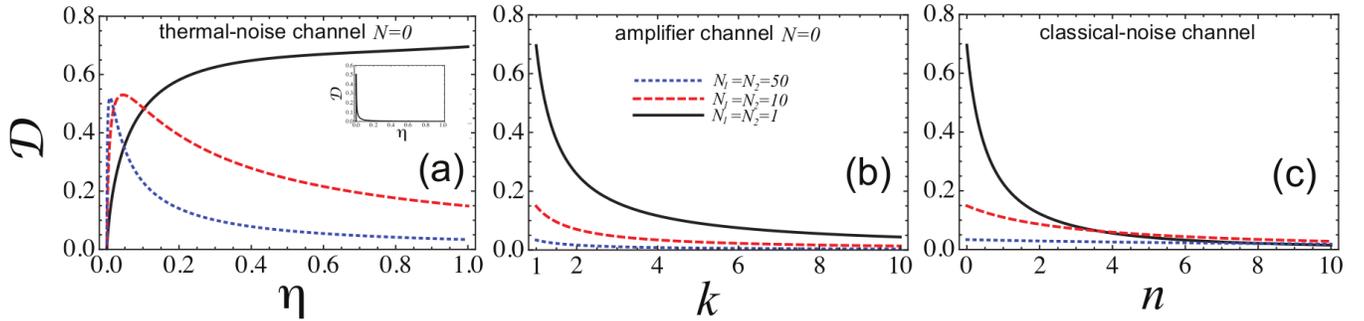}
\caption{(Color online){ Behavior of Gaussian discord $\mathcal{D}$ under a thermal-noise (a), an amplifier (b) and a classical-noise channel (c) {with the former two taken for $N\ug0$}. Throughout, we have considered an initial state (\ref{sts}) for $r\ug1$ and $N_1\ug N_2\ug1$ (solid black lines), $N_1\ug N_2\ug10$ (red dashed) and $N_1\ug N_2\ug100$ (blue dotted). The inset in (a) shows  $\mathcal{D}$ against $\eta$ for $N_1\ug N_2\ug1000$ under the zero-temperature thermal-noise channel.}
 \label{Fig1}}
\end{figure*} 
To begin with, we consider initial STSs [\cf\eqs(\ref{sts}) and (\ref{thermal})] having $r\ug1$ and $N_1\ug N_2$. In \fig1, we plot the behavior of Gaussian discord (\ref{disc}) for increasing values of $N_1\ug N_2$ in the presence of the thermal-noise and amplifier channel (both with $N\ug0$) as well as the classical-noise channel. We recall that each local non-unitary map affects only mode 2.
The classical-noise channel [see \fig1(c)] has a merely detrimental effect on the QCs, which monotonically decay for a growing number of noise photons regardless of $N_1$. A similar effect is exhibited in the case of the amplifier channel when the gain parameter $k$ is increased [see \fig1(b)]. In  contrast, however, a non-monotonic behavior can take place with the lossy channel, as shown in \fig1(a). For low average photon numbers of each mode $N_1\ug N_2$, similarly to \figs1(b) and (c), a monotonic drop of QCs occurs as the quantum efficiency $\eta$  decreases (we recall that the corresponding map reduces to the identity operator for $\eta\ug1$). As the photon number becomes larger, though, the discord remarkably undergoes a {\it rise} so as to reach a maximum value and eventually drop to zero. Such increase can be quite significant. For instance, as shown in \fig1(a), in the case $N_1\ug N_2\ug10$ the maximum taken by $\mathcal{D}$ is about $2.5$ times the initial value (for $\eta\ug1$) and even $\simeq\!10$ larger when $N_1\ug N_2\ug50$. For higher photons numbers (still keeping fixed the set value of $r$) the initial discord becomes extremely low but a significant rise still takes place before the asymptotic decay. We illustrate the latter feature in the inset of \fig1(a) for the paradigmatic case $N_1\ug N_2\ug1000$, which corresponds to a fully separable state \cite{nota-sep} with an initial discord $\mathcal{D}|_{\eta=1}\!\simeq\!10^{-6}$. Yet, under the lossy channel this undergoes a rise reaching a maximum $\simeq\!0.5$ before the eventual slow decay. 
\begin{figure}
 \includegraphics[width=0.4 \textwidth]{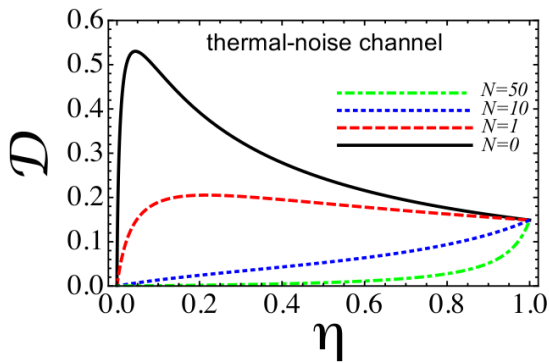}
\caption{(Color online){ Gaussian discord $\mathcal{D}$ against $\eta$ under the thermal-noise channel for the average photon number of the reservoir $N\ug0$ (lossy channel, solid black line), $N\ug1$ (red dashed), $N\ug10$  (blue dotted) and $N\ug50$  (green dot-dashed). Throughout, the considered initial state is (\ref{sts}) with $r\ug1$ and $N_1\ug N_2\ug10$.}}
\end{figure} 
For all practical purposes, one can thus regard this behavior as mere creation of (previously absent) quantum correlations, which is significantly reminiscent of an analogous effect occurring for qubits under local amplitude-damping channels \cite{ciccarello, mauro,bruss}.

It is natural to wonder how the discord rise is affected in the presence of a reservoir having non-zero temperature. To investigate this, in \fig2 we display $\mathcal{D}$ against $\eta$ in the case of the thermal-noise channel for growing values of the reservoir average photon number $N$. Clearly, the effect of temperature is to spoil the discord increase. At high enough temperatures, the initial rise is no more exhibited and the behavior reduces to a mere monotonic decay similarly to the amplifier and classical-noise channels [\cf \figs1(b) and (c)]. {Note that there is a threshold value for $N$ separating the non-monotonic regime (featuring QC's rise) from the monotonic one, the latter occurring above the threshold. For set $r$, such critical value linearly grows with $N_1$. This is shown in \fig3, where we plot the initial slope of the Gaussian discord $p$, \ie $p\ug\partial \mathcal{D}/\partial \eta$ at $\eta\ug0$, \vs $N$ and $N_1\ug N_2$. By recalling that $\eta$ decreases during the system's evolution under the thermal-noise channel, $p\!>\!0$ ($p\!<\!0$) means that an initial decay (rise) of QCs occurs. Hence, the intersection of $p(N,N_1)$ with with the $N\!-\!N_1$ plane provides the functional dependance of the aforementioned threshold value on $N_1$.}
\begin{figure}
 \includegraphics[width=0.4 \textwidth]{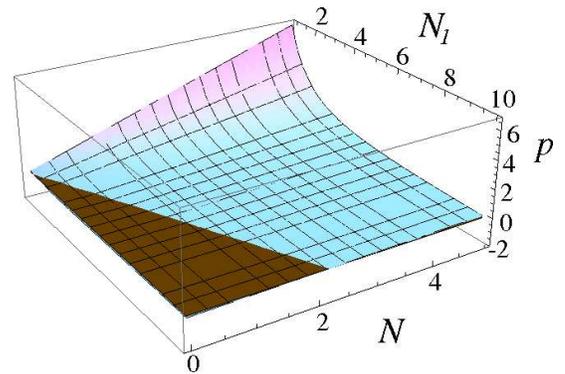}
\caption{(Color online){ Initial slope of the Gaussian discord $p$, \ie $\partial \mathcal{D}/\partial \eta$ at $\eta\ug0$, \vs
$N$ and $N_1\ug N_2$ $\eta$  for $r\ug1$ under the thermal-noise channel. The intersection line between the function profile (in blue) and the plane $p\ug0$ (in brown) specifies, for set $N_1$, the maximum value of $N$ yielding QCs' rise.}}
\end{figure} 
\begin{figure*}
 \includegraphics[width=0.95\textwidth]{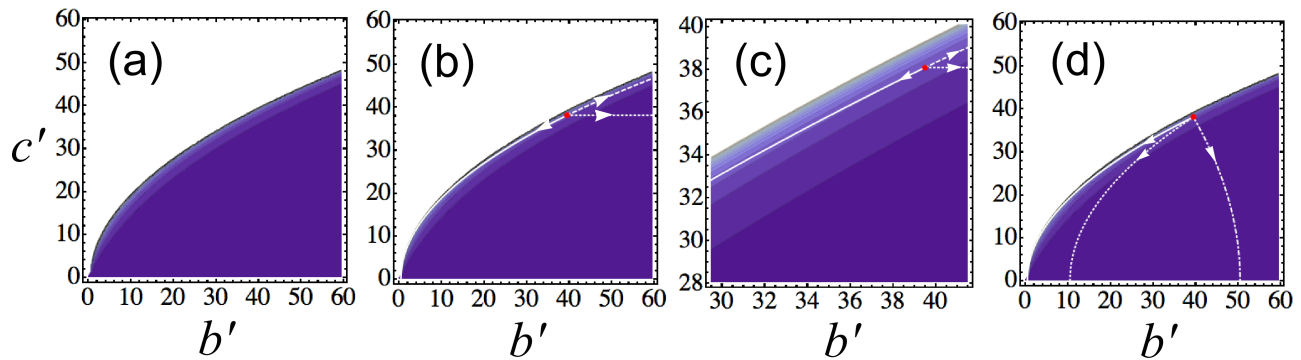}
\caption{(Color online){ (a) Contour plot of Gaussian discord $\mathcal{D}$ against $b'$ and $c'$. (b) Oriented trajectories in the $b'\!-\!c'$ plane arising with
 the lossy channel (solid line), the zero-temperature amplifier channel (dashed) and the classical-noise channel (dotted). The red point corresponds to the initial state. (c) Zoom of the previous figure highlighting the initial rise of discord occurring only under the lossy channel. (d) Effect of temperature in the case of the thermal-noise channel: oriented trajectories for $N\ug0$ (lossy channel, solid line), $N\ug1$ (dashed), $N\ug10$ (dotted) and $N\ug50$ (dot-dashed).
  Throughout, we have considered an initial state (\ref{sts}) for $r\ug1$ and $N_1\ug N_2\ug10$.}}
\end{figure*} 

\section{Insight into the rise of quantum correlations} \label{insight}

Here, we give a picture that illustrates the emergence of some key features and behaviors presented in the previous Section. 
We focus on the initial state specified by $r\ug1$ and $N_1\ug N_2\ug10$ exhibiting a rise of discord followed by a decay in the case of the thermal-noise channel (at low temperatures) and a monotonic drop with the other two channels (\cf\figs1 and 2). As in each of such three cases the environment does not affect parameter $a'$ [see \eqs(\ref{abctnc}), (\ref{abcac}) and (\ref{abccn})] and the equality $c'_1\ug-c'_2$ always holds, in the addressed regime the discord is in fact a function of $b'$ and $c'\ug|c_1'|\ug|c_2'|$ [\cf \eq(\ref{disc})]. As shown in the contour plot in \fig4(a), in its domain of definition $\mathcal{D}$ decreases with $b'$ and increases with $c'$. One can now obtain insight into the behavior of QCs by observing that the equations for $b'$ and $c'$ define a characteristic curve (associated with the specific channel) parametrized by $\eta\!\in\![0,1]$, $k\!>\!1$ and $n\!>\!0$ in the case of the thermal-noise, amplifier and classical-noise channel, respectively. By expressing each of such parameters as a function of $c'$ and replacing it into the equation for $b'$, one obtains the trajectory of each channel in the $b'\!-\!c'$ plane as
\begin{eqnarray}\label{trajectories}
b'&\ug& \frac{b\meno \left(N\piu1/2\right)}{c^2}c'^2+(N\piu1/2)\,\,\,\,\,\,\,\,\,\,\,\left(c'\in[0,c]\right)\,\,,\label{tratn}\\
b'&\ug& \frac{b\piu \left(N\piu1/2\right)}{c^2}c'^2-(N\piu1/2)\,\,\,\,\,\,\,\,\,\,\,\left(c'\in[c,\infty]\right)\,\,,\label{traac}\\
c'&\ug& c \,\,\,\,\,\,\,\,\,\,\,\left(b'\!\in\![b,\infty]\right)
\end{eqnarray}
In \figs4(b) and (c), we report the above trajectories along with the same contour plot of discord as in \fig4(a). Evidently, due to the above discussed functional shape of $\mathcal{D}$ a monotonic decrease of QCs necessarily takes place under the amplifier and classical-noise channels. It is also clear from the bottom left portion of \fig4(b) that for the lossy channel as $\eta$ is decreased $\mathcal{D}$ must eventually drop. However, the trajectory is such that over the first stage ($\eta$ slightly below 1) the QCs {\it grow} [this is best illustrated by {the zoom presented} in \fig4(c)].

The effect of temperature in the case of the thermal-noise channel (\cf \fig2) can be understood by scrutinizing \eq(\ref{tratn}) and \fig4(d). For $\eta\!\rightarrow\!0$ $c'$ tends to zero and thereby $b'\!\rightarrow\!(N\piu1/2)$. Hence, the higher $N$ (\ie the temperature) the larger the asymptotic value of $b'$. Accordingly, the concavity $C\ug b\meno \left(N\piu1/2\right)/c^2$ -- which is positive for $N\ug0$ -- progressively decreases so as to eventually become negative. As soon as it becomes small enough, the shape of $\mathcal{D}(b',c')$ [\cf\fig4(a)] prevents from any initial rise to take place, which results in a monotonic decay (see \fig2).

\section{conclusions} \label{concl}

In this paper, we have investigated the behavior of Gaussian discord for a two-mode squeezed thermal state subject to various local Gaussian channels. Mainly motivated by the finding that a local amplitude-damping channel can create QCs in a pair of qubits, we have explored the dynamics under the thermal-noise, amplifier and classical-noise channel. While the typical behavior is a monotonic decrease of QCs, {as one would expect}, we have found that in significant analogy with the qubits' framework a thermal-noise channel can give rise to a non-monotonic behavior comprising an initial {\it rise} of discord. {For large enough photon numbers of each mode such that the initial QCs are in fact negligible, the above entails creation of previously absent QCs for all practical purposes.}
The reservoir temperature spoils this phenomenon in a way that when it is high enough a mere monotonic decay occurs.

We have provided a picture that allows to shed light on various features, in particular the reason why the discord rise is exhibited only under the thermal-noise channel as well as the detrimental effect of temperature on it.

These findings significantly extend to the CV-variable scenario one of the most {counter-intuitive} effects entailed by the emerging extended paradigm of QCs: {quantum correlations can be established in a composite system through the interaction with a local and memoryless bath.} 

\begin{acknowledgments}
{We thank Matteo Paris and Mauro Paternostro for comments and acknowledge support from FIRB IDEAS through project RBID08B3FM.}
\end{acknowledgments}

\end{document}